\documentclass[twocolumn,showpacs,preprintnumbers,amsmath,amssymb,bbold]{revtex4}
\usepackage{epsfig,amsopn}
\usepackage{graphicx}
\usepackage{epstopdf}
\usepackage{sidecap}
\usepackage{color}
\usepackage{amsmath,amssymb}
\usepackage{amsthm}
\usepackage{enumerate}

\usepackage{setspace}
\newcommand\bea{\begin{eqnarray}}
\newcommand\eea{\end{eqnarray}}
\newcommand\beq{\begin{equation}}  
\newcommand\eeq{\end{equation}}
\newcommand{\new}{\newpage}

\begin{document}
\title{Bulk-edge correspondence and topological phases in periodically driven spin-orbit coupled materials in the low frequency limit}
\author{Ruchi Saxena$^{1}$, Sumathi Rao$^{2}$ and Arijit Kundu$^{3}$}
\affiliation{\mbox{$^1$} {Physics Department, Advanced Technology Institute, University of Surrey, Stag Hill, Guildford, GU2 7XH, United Kingdom} \\
\mbox{$^2$} {Harish-Chandra Research Institute, HBNI, Chhatnag Road,   Jhunsi, Allahabad 211 019, India} \\
\mbox{$^3$}{Department of Physics, Indian Institute of Technology - Kanpur, Kanpur 208 016, India}}
\begin{abstract}

We study the topological phase transitions induced in spin-orbit coupled materials with buckling like silicene, germanene,
stanene, etc,  by circularly polarised light, beyond the high
frequency regime, and unearth many additional topological phases.
We also study the robustness of these phases in the presence of uniform disorder.
These phases are characterised by the spin-resolved
topological invariants, $C_0^\uparrow$, $C_0^\downarrow$, $C_\pi^\uparrow$ and $C_\pi^\downarrow$, which specify
the spin-resolved edge states traversing the gaps at zero quasi-energy and the Floquet zone boundaries respectively. We show that for each phase boundary, and independently  for each  spin sector,  the gap closure in the Brillouin zone occurs at  a high symmetry point. 

\end{abstract} 
\maketitle

\section{Introduction}  

Dynamical control of topological phases is one of the most intensely researched topics in recent times\cite{Oka2009,Kitagawa2010,Gu2011,Lindner2011,Jiang2011, Calvo2011,Dora2012,Kundu2013,Kundu2014,Ezawa2013}.
Proposals have involved periodic driving in semiconductor systems\cite{Lindner2011}, cold atom  (or optical lattice) systems\cite{Jiang2011}, graphene\cite{Oka2009,Kitagawa2010,Gu2011,Kundu2014} and 
systems with spin-orbit coupling like silicene\cite{Ezawa2013},  with a variety of analytical and numerical methods. Apart from band-structure control of a system by renormalization of its dynamical parameters via a periodic drive, novel  non-trivial topological phases, which do not have  any analog in static systems, have  been explored, theoretically ~\cite{Rudner2013,Platero2013,Platero2014,Carpentier2014,Sentef2015,Fistul2015}
as well as in experimental photonic systems\cite{Maczewsky2017,Mukherjee2017expt}.

Silicene~\cite{Ezawa2011a}, and other spin-orbit 
coupled~\cite{Konschuh2010} materials like germanene, stanene, etc  are 
recently synthesized materials which have shot into prominence because 
their buckled nature allows them to be tuned by an electric field through 
a transition between a band insulator and a topological 
insulator~\cite{Kane2005,Hasan2010,Qi2011}. 
This tunability, and in particular, the experimental realisation~\cite{Tao2015}  of silicene-based transistors has led 
to extensive 
work~\cite{Linder2014,Rachel2014,Saxena2015,Paul2016,Li2016,Sarkar2016}  
on the interplay of topology and transport in these materials. 

More recently, the question of whether topological phases can be 
controlled in irradiated silicene and similar materials has been studied. 
Although these materials are time-reversal invariant, in the 
presence of circularly polarised light, the system breaks time-reversal 
symmetry and the Chern number classification is integer and not $Z_2$ 
invariant. Ezawa~\cite{Ezawa2013}  showed that 
at high frequencies and for small amplitudes of driving, new phases 
in silicene such as quantum Hall insulator, spin-polarised quantum Hall 
insulator and spin and spin-valley polarised metals can be realised. 
Further, it was shown~\cite{Mohan2016}  that many more topological phases 
could be realised by performing a systematic Brillouin-Wigner (BW) 
expansion of the Hamiltonian to second order in the inverse of the 
frequency, not only in silicene, but in other spin-orbit coupled 
materials. But, as was discussed also in earlier 
references~\cite{Mohan2016,Mikami2016}, the BW expansion breaks down when 
the frequency $\omega$ becomes smaller than the band-width of the 
effective Hamiltonian. The real constraint on the applicability of the BW 
expansion is a combined bound on both $\omega$ and the amplitude of 
driving and in fact, the validity of BW increases, even for lower 
frequencies  when the amplitude increases. However, the physical reason 
for the breakdown of the earlier studies at low frequencies is because,  
at frequencies  comparable to the band-width,  it is no longer possible to 
neglect the topology of the \textit{quasienergy} space, which forms a 
periodic structure with the single valuedness of the  eigenfunction 
requiring  the quasienergies to be within a ``Floquet zone''.  Low 
frequency driving can lead to  crossings between the bottom of one Floquet 
band and  the top of the next  Floquet band. These crossings are neglected 
in the BW expansion and hence, the study of the driving at low frequencies 
requires a new formalism which goes beyond the effective static 
approximation of a dynamical Hamiltonian.

To characterize the topological nature of these system that would 
satisfy the edge-bulk correspondence, one needs to have access to the  full 
time-dependent bulk evolution operator $U(t)$, evaluated for all 
intermediate times within the driving period~\cite{Rudner2013}. The 
invariants thus computed predict the complete Floquet edge-state spectrum. 
Similar $Z_2$ valued indices for periodically driven time-reversal 
invariant two dimensional indices have also been 
found~\cite{Carpentier2014}. For a two band model with the Fermi energy 
at zero quasi-energy,  it was shown that the  gaps at zero quasi-energy 
and at the zone boundary  $\omega/2$ gave rise
to winding numbers  $C_0$ and $C_\pi$, whose difference gave the Chern 
number of the band. This formalism has been used to 
demonstrate the bulk-boundary correspondence in graphene 
~\cite{Kundu2014, graphene}. 

However, most of the studies have focused on graphene both in the high 
and low frequency regimes  and silicene-like materials have not been 
fully explored when they are exposed to low frequency radiation. In 
this paper, we study the photo induced topological phase diagram in 
these materials by computing the topological invariant for both up and 
down spin. We also look for the gap closing points in the Brillouin 
zone and find that it always occur at the high symmetry points.  
Further, we compute $C_0$ and $C_\pi$ by counting the number of edge 
states at the right and left edges of the sample and verify the 
bulk-boundary correspondence as illustrated earlier in graphene ~\cite{Kundu2014,graphene}. 
Although the question of  whether these 
topological phases obtained in the low frequency regime are stable or not, 
requires detailed analysis to answer definitively, here 
%which requires detailed analysis to answer 
we   just address the behaviour of a few phases 
against uniform disorder by computing the real space Chern number 
using the coupling matrix approach prescribed in ~\cite{coupling}. 
A full analysis is left for future studies. We further note 
that our study can also be extended to two dimensional optical lattices 
where one can artificially synthesise an effective spin-orbit coupling by a 
combination of microwave driving and
lattice shaking ~\cite{optical},  and also to ultra cold atoms 
~\cite{coldatom}, in both bosonic ~\cite{bosonic_1,bosonic_2}
and fermionic systems ~\cite{fermionic1,fermionic2} where the 
Raman coupling has been shown to give rise to an effective spin-orbit 
interaction. 
 
%--------------- Fig 1 -------------------
 \begin{figure}
 \begin{center}
\hspace*{-0.5cm}\includegraphics[width=0.55\textwidth]{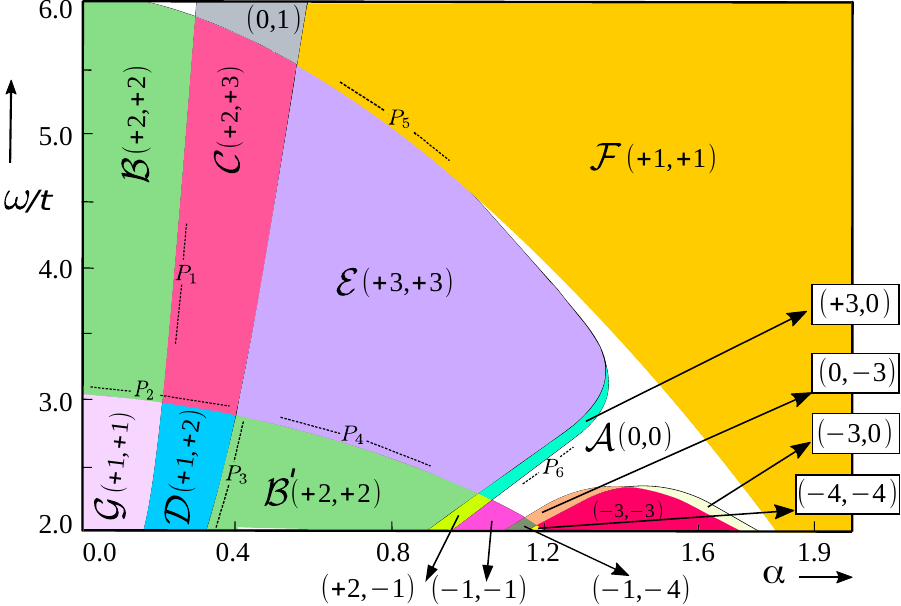}
   \end{center} 
   \vspace*{-0.2cm}
  \caption{The phase diagram for the model Hamiltonian in Eq.~\ref{hamil}  
as we vary the amplitude $\alpha$ and the drive frequency $\omega$,  with 
the    external electric field  fixed at $lE_z=0.08 t$ and spin-orbit 
coupling, $\lambda_{SO}$ at 0.05t. Each phase is characterized by the 
spin resolved quantum numbers ($C_\uparrow$, $C_\downarrow$). We label 
the phases by calligraphic letters. The dotted lines $P_i$ indicate the 
topological phase boundaries, inferred from the gap closing in  momentum 
space, shown only for the $\uparrow$ spin sector. }
  \label{figone}
 \end{figure}
%-----------------------------------------

\section{Computation of the dynamical  band structure}
We start with two  dimensional Dirac systems which are buckled due to the large ionic radius of the silicon atoms and consequently have a non-coplanar structure unlike graphene.  These materials can be described by a four-band tight binding model in a hexagonal lattice given by 
\begin{align}
H = -t \sum_{\langle i,j\rangle,\sigma} c_{i\sigma}^\dagger c_{j\sigma}+\frac{i \lambda}{3 \sqrt{3}} 
\sum_{\langle \langle i,j\rangle\rangle,\sigma} \sigma \nu_{i,j} c_{i \sigma}^\dagger c_{j\sigma} \nonumber \\
  + l E_z\sum_{i \sigma} \xi_i c_{i \sigma}^\dagger c_{i\sigma}~. \label{hamil}
\end{align}
Here, the first term is the kinetic term where $t$ is the hopping parameter. The second term represents the spin-orbit coupling term where the value of $\lambda$ depends on the material and $\nu_{i,j} = \pm 1$ depending on whether the next-nearest 
neighbour hopping is clock-wise or anti-clock-wise. The last term represents the staggered sub-lattice potential due
to the buckling. When a beam of circularly polarised light 
%(\textcolor{red}{explain why we use circularly polarised light and how things will
%change if we use linearly or elliptically polarised light}) 
is incident on the sheet, the corresponding electro-magnetic potential
${\bf A} = (A_0\cos(\omega \tau), A_0\sin(\omega \tau), 0 )$ is introduced into the Hamiltonian using Peierls substitution. $\omega$ is the frequency of light and $A_{0}$ is  its amplitude. In the Fourier transformed space, this is written as
\begin{align}
H(\tau)=\left(
    \begin{array}{cccc}
     l E_z -\delta_\lambda & \delta_t & 0 & 0 \\
    \delta_t^* & -l E_z+\delta_\lambda & 0 & 0   \\
    0 & 0 & l E_z+\delta_\lambda & \delta_t \\
    0 & 0 & \delta_t^* & -l E_z-\delta_\lambda 
    \end{array} 
    \right)  
\end{align}
where 
 \begin{align}
\delta_\lambda(\tau) = \frac{2  \lambda}{3 \sqrt{3}} \left[ \sqrt{3}a_0 \sin  {\tilde k_x} -\sin \left (
\frac{\sqrt{3}a_0}{2}{\tilde k_x} +  \frac{3 a_0}{2}{\tilde k_y} \right) \right.&\nonumber   \\
\left. - \sin\left(\frac{\sqrt{3}a_0}{2}{\tilde k_x} - \frac{3a_0}{2}{\tilde k_y}\right) \right] & 
 \end{align}
 with ${\tilde k_x} = k_x+A\cos\omega\tau$ and ${\tilde k_y} =  k_y+A\sin\omega\tau$  and 
\begin{align}
 \delta_t(\tau) &= t \left[ {\rm exp}(-i \alpha\sin \omega \tau) \right.\nonumber \\
 &\left.+ T_+ {\rm exp}\frac{i \alpha( \sqrt{3}\cos \omega  \tau + \sin \omega \tau)}{2} \right . \nonumber \\
&\left. + T_- {\rm exp}\frac{i \alpha(-\sqrt{3}\cos \omega  \tau + \sin \omega  \tau)}{2}  \right]
 \end{align}
with $T_\pm = {\rm exp} (i a_0(\pm \sqrt{3}k_x+{3k_y}/2))$. Here, we have defined $\alpha = Aa_0$, where $a_0$ is the lattice constant.
 
For the bulk system, the vector potential and hence the Hamiltonian  is periodic in both the $x$ and $y$ directions.
This implies that we can rewrite the Hamiltonian in  terms of a Floquet eigenvalue problem with
the Hamiltonian given by 
\begin{align}
H_F  = -i \frac {\partial } {\partial \tau} + H(\tau), \label{ev}
\end{align}
the eigen functions given by 
\beq
\psi_{{\bf k},b} (x,y, \tau) = u_b(k_x,k_y,\tau)e^{i{\bf r}\cdot {\bf k} - i\epsilon_b \tau} \label{efn}
\eeq 
with $u_b(k_x,k_y,\tau)
=u_b(k_x,k_y,\tau+2\pi/\omega)$, 
and where $\epsilon_b$ are  the quasienergies or the eigenvalues of $H_F$. 
The Hamiltonian can now be solved numerically
as a function of the amplitude $A_0$, frequency $\omega$ and the sub-lattice potential $E_z$, both for the quasienergy
eigenvalues and for the wave-functions.

At high frequencies,   $\omega$ constitutes a large gap between unperturbed subspaces,
and the extended Floquet Hilbert space splits into decoupled subspaces with different photon numbers.  
Since the perturbation scale of the Hamiltonian,  which is the band-width $t$,  is much smaller than $\omega$, one can use systematic perturbation theory to include virtual processes of emitting and absorbing photons,  and 
 upto a given order in perturbation theory,  one can obtain an effectively static Hamiltonian as shown
 in Ref.~\onlinecite{Mohan2016}. The Chern numbers for the model can then be computed by integrating the Berry curvature over the whole Brillouin zone \cite{Hatsugai_2005} using the 
 eigenvectors of the effective Hamiltonian. However it is expected that such an expansion in $1/\omega$ would fail to predict the correct Chern numbers once the frequency of the drive, $\omega$, becomes comparable to the bandwidth. This is the part of the phase diagram that we shall complete in this paper.
 
 %--------------- Fig 2 -------------------
 \begin{figure}
 \begin{center}
\includegraphics[width=0.48\textwidth]{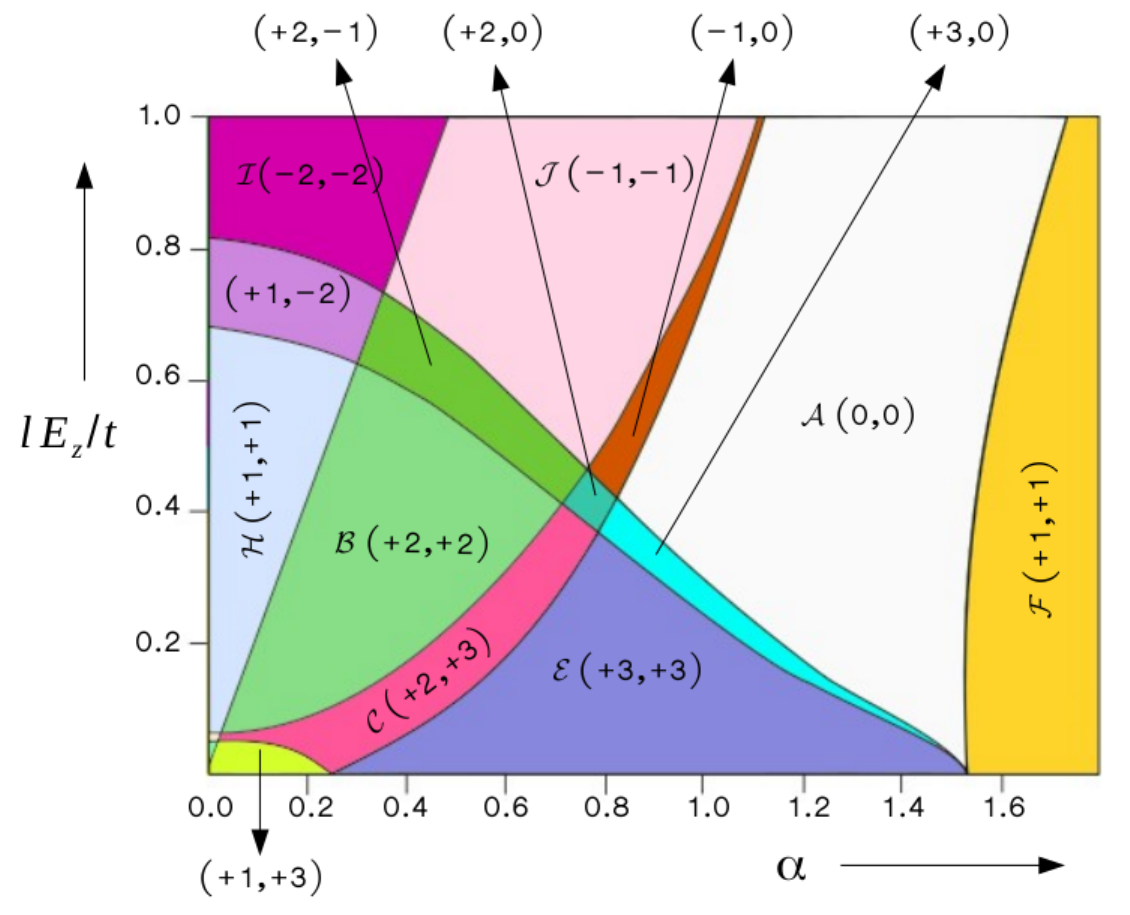}
   \end{center} 
   \caption{The phase diagram as a function of the amplitude $\alpha$ and the external electric field $lE_z$. A low  frequency drive  is chosen ($\omega=3.0 t$) since we wish to  study the system in the   low frequency
   limit. All the phases are the same as those found in Fig.\ref{figone} except for three new phases - $\mathcal{H}$, $\mathcal{I}$ and  $\mathcal{J}$. The labelling of the phases follows the same convention as in Fig.~\ref{figone}.}  \label{figtwo}
 \end{figure}
%-----------------------------------------

\subsection{The phase diagram of the Floquet Hamiltonian}
As the frequency of the drive becomes comparable to the effective bandwidth of the system, it is essential to now consider the complete nature of 
the quasi-energy bands in the computation of the topological invariants of the system. As was mentioned in the introduction, the quasi-energy bands
(of the two band system) are now identified with two topological invariants, $C_0$ and $C_{\pi}$ and the net Chern number of a band is given by $C = C_0 - C_{\pi}$ (independently for each of the spins).

The Fourier-transformed time-dependent Hamiltonian (Eq.~2) is block-diagonal in the spin space. For either the  $\uparrow$ or the $\downarrow$ spin, it is a 2 $\times$ 2 Hermitian matrix which 
encodes the bulk properties of the system. The time evolution operator at stroboscopic times can then be written as
\begin{align}
U(\textbf{k}, 2\pi/\omega) = \mathcal{T} e^{-i \int_{0}^{2\pi/\omega} H(\textbf{k},\tau) d \tau}. \label{oper}
\end{align}
and the Floquet states $u_b(k_x,k_y,0)$ are the eigenstates of this  operator. The Chern number of each Floquet band is then defined by integrating the Berry curvature of the Floquet states over the whole Brillouin zone - 
\begin{align}
C = \frac{1}{2\pi} \int_{BZ}^{} dk_x dk_y (\nabla \times \mathcal{A}_{\text{lower}}(\mathbf{k})),
\end{align}
where $\mathcal{A}_{\text{lower}}$ is the Berry connection in terms of Floquet states of the quasi-energy band with quasienergy lying between ($-\omega/2$, 0). We numerically compute the Chern numbers of the lower band (of both $\uparrow$ and $\downarrow$
spins) following the work by Fukui et al~\cite{Hatsugai_2005}.

When the parameter ranges are such that a high frequency approximation would be valid, the Chern number computed using the effective static Hamiltonian would exactly match the one obtained by considering the Floquet states.  In this sense, the following phase diagram that we present complements what has been obtained earlier in  Ref.~\onlinecite{Mohan2016}, and  completely specifies the topological phases of the system for all parameter regimes.

%--------------- Fig 3 -------------------
\begin{figure}
	\begin{center}
		\includegraphics[width=0.35\textwidth]{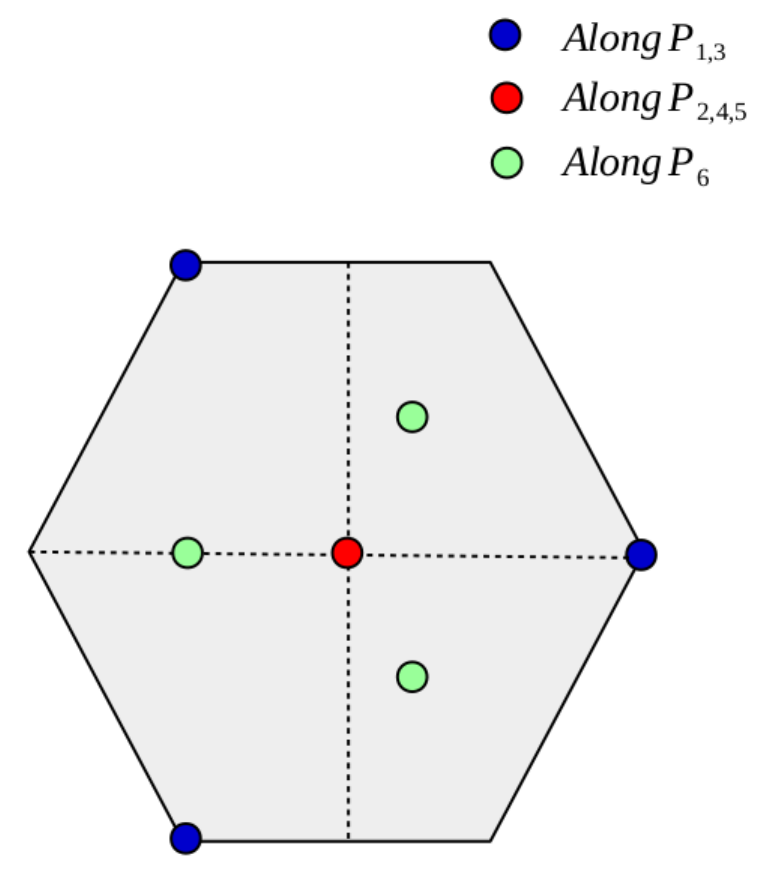}
	\end{center} 
	\caption{Gap closing points in the Brillouin zone along the $P_i \,\, (i=1\dots 6)$  phase boundaries (drawn in Fig.1) as described in the caption of Fig.1. } 
	\label{figthree}
\end{figure}
%-----------------------------------------

The  phase diagrams for both the up spin  and the down spin bands are presented in Figs.~\ref{figone} and \ref{figtwo}. In Fig.~1, we show the Chern number of the lower quasienergy band as a function of the amplitude of the drive versus the frequency, whereas in Fig.~2 we show it as a function of the amplitude of the drive versus the sub-lattice potential.  For lower frequencies, many different phases appear and appear  to follow a fractal structure, as was seen for graphene in Ref.\cite{Mikami2016}. But as such phases are not expected to be protected by a large enough band-gap, we have only shown phases which are `large enough' (occupy enough area in the phase diagram) and we have ignored tinier phases. As $\alpha \rightarrow 2$ and $\omega \rightarrow 6$, these phases smoothly go over to the high frequency phases in Ref.~\onlinecite{Mohan2016}.  We have also chosen to name only those phases that are large enough to be possible stable phases in calligraphic letters as $\mathcal{A}, \mathcal{B} \dots \mathcal{J}$, with $\mathcal{A},\mathcal{B}, \mathcal{C},\mathcal{E},\mathcal{F}$ being present in both Figs.~\ref{figone} and \ref{figtwo}, and $\mathcal{B'},\mathcal{D},\mathcal{G}$ in Fig.\ref{figone} and $\mathcal{H}, \mathcal{I},\mathcal{J}$ in Fig.~\ref{figtwo}.  Note that there are two phases $\mathcal{B}$ and $\mathcal{B'}$ which have identical  values of the Chern numbers for both the $\uparrow$ spin band and the $\downarrow$ spin band. Nevertheless, they are two distinct phases since they occur for  different values of $\omega$ and $\alpha$ and are not continuously connected to each other and they could have different edge state structures. Note also the existence of a phase $\mathcal{A}$ which has zero Chern numbers for both spin $\uparrow$ and spin $\downarrow$ electrons. We will see later in the next section, that this is a topological phase and has edge states despite having zero Chern numbers.

The lines that separate the phases are when the gap closes and the gap closing typically occurs at the  high symmetry points of the Brillouin zone as shown in Fig.~\ref{figthree}.  For the lines $P_2,P_4$ and $P_5$, the gap closes at the $\Gamma$ point  whereas for  the $P_1$ and $P_3$ lines,
it closes at the $K$ point and for the  $P_6$ line, the closure happens at the half-way point between the $\Gamma$ point and the $K$ point. Note that we have concentrated on the spin $\uparrow$ bands and hence have lines separating region $\mathcal{C} $ from $\mathcal{E}$,  which have different Chern numbers for
$\uparrow$ spin, but no line separating regions $\mathcal{C}$ from $\mathcal{B}$, which have the same Chern number for $\uparrow$ spin. A similar analysis can be done for the $\downarrow$ spin case.

We note that the Chern number changes by $\pm 2$ at the $P_5$ crossing, which essentially implies a
quadratic touching of the bands.  This is similar to the transition explained in Ref.~\onlinecite{Kundu2014} where the Hamiltonian for the first $\Gamma$ point transition at the Floquet zone boundary was  obtained perturbatively, and was shown to lead to a Chern number change of $\pm 2$.  This can only happen at the spherically symmetric $\Gamma$ point. Along $P_1,P_2,P_3$ and $P_4$, the change in the Chern number is $\pm 1$ and the band touching happens at the $\Gamma$ or $K$ points.  Along $P_6$, however, the change in the Chern number is $\pm 3$. This can happen at 3 points in the Brillouin zone, symmetric around the $\Gamma$ point as shown in Fig.~\ref{figthree}.  We have also checked that a change of the chirality of the circularly polarized light, besides changing signs of all the Chern numbers also breaks inversion symmetry with respect to the gap closing diagram in Fig.~\ref{figthree}.  The blue points are at $K'$ instead of $K$ points and the green points are placed so as to complete the smaller hexagon.

%--------------- Fig 4 -------------------
\begin{figure}
	\begin{center}
		\includegraphics[width=0.48\textwidth]{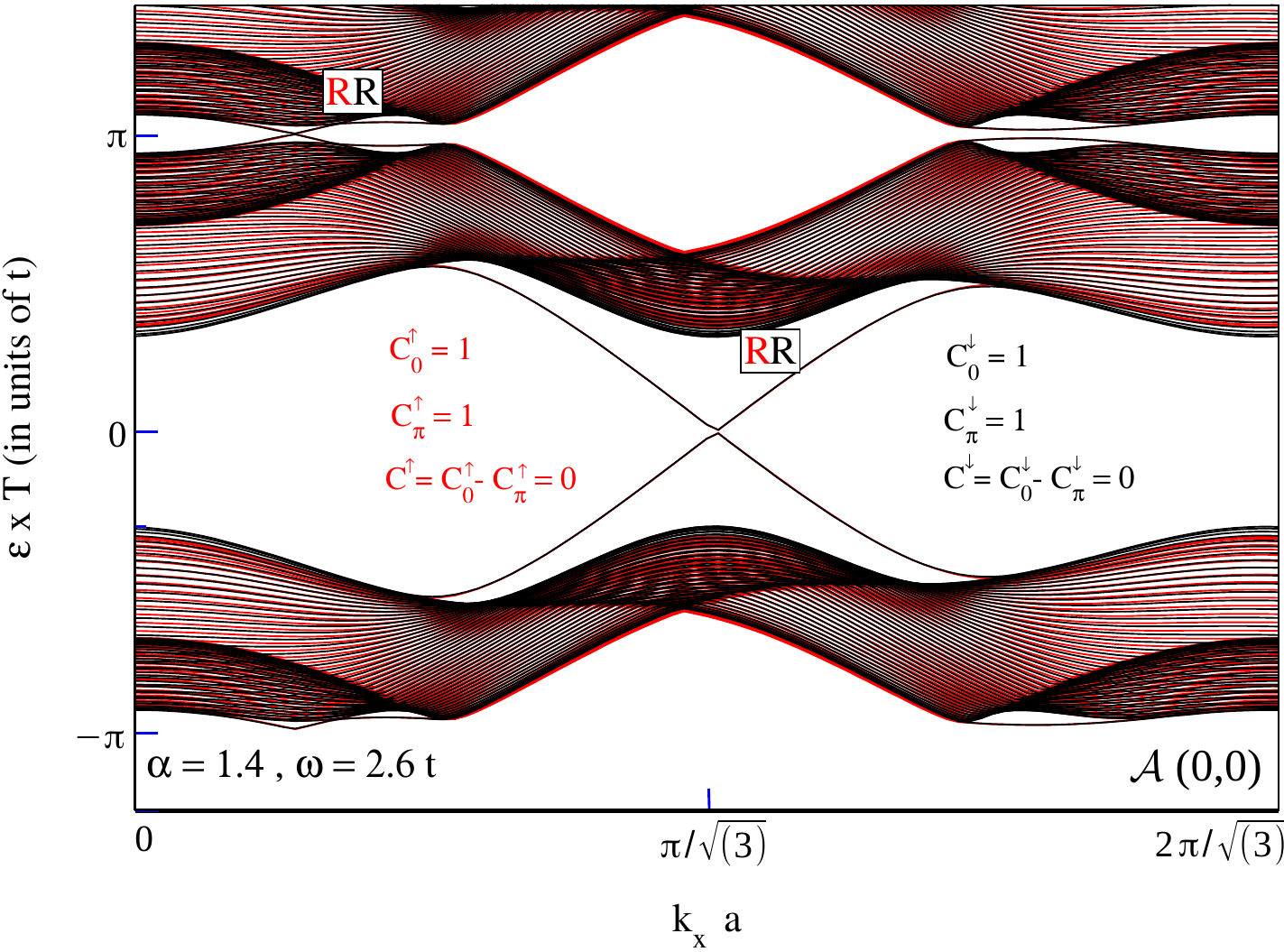}
	\end{center} 
	\caption{The quasi-energy band structure of a zigzag nanoribbon of the periodically driven spin-orbit coupled system for phase $\mathcal{A}$. Both spin sectors, $\uparrow$ and $\downarrow$
	 (shown in red and black respectively) possesses one pair 
	of chiral edge states both at zero quasi energy and Floquet zone boundary. We also label the chirality of  the  left edge state at the two inequivalent gaps by R or L depending on  whether the state is right-moving or left-moving. The
	system is finite in the $y$-direction while the $x$-direction is periodic. } 
	\label{figfour}
\end{figure}
%-----------------------------------------
However, the computation of the Chern number does not specify the $C_0$ and $C_{\pi}$ invariants individually. As the bulk-boundary correspondence in our system comes from these invariants, to discover these two indices, we  need to consider the edge-state structure in a system with edges - $e.g$., a ribbon geometry. This is what we shall discuss in the following section.

\begin{table}[]
	\centering
	\caption{Spin-resolved topological quantum numbers and the edge states for phases in Figs.\ref{figone}, \ref{figtwo}}.
	\label{tabone}
	\begin{tabular}{|c|c|c|c|c|c|} \hline
		Phases &~~~($C^\uparrow, C^\downarrow$)~~~ & ~~~$C_0^\uparrow$ ~~~ & ~~~$C_\pi^\uparrow$ ~~~& ~~~$C_0^\downarrow$ ~~~& ~~~ $C_\pi^\downarrow$~~~ \\ \hline
		$\mathcal{A}$ & (0,0) &  1& 1 & 1 & 1\\ \hline
		$\mathcal{B},\mathcal{B'}$ & (+2,+2) & 0  & $-2$  & 0  &  $-2$ \\ \hline
		$\mathcal{C}$ & (+2,+3) & 0 & $-2 $ & 1  & $-2$ \\ \hline
		$\mathcal{D}$ & (+1,+2) & $-1$ & $-2$  & 0  & $-2$ \\ \hline
		$\mathcal{E}$ & (+3,+3) & 1 & $-2$ & 1 & $-2$\\ \hline
		$\mathcal{F}$ & (+1,+1) & 1  & 0 & 1 & 0\\ \hline
		$\mathcal{G}$ & (+1,+1) & $-1$ & $-2$ & $-1$ & $-2$\\ \hline
		$\mathcal{H}$ & (+1,+1) & 0 & $-1$ & 0 & $-1$ \\ \hline
		$\mathcal{I}$ & ($-2,-2$) & 0 & 2 & 0 & 2 \\ \hline
		$\mathcal{J}$ & ($-1,-1$) & 0 & 1 & 0 & 1 \\ \hline
	\end{tabular}
\end{table}

\subsection{Edge states in a ribbon geometry}

In this section, we study the quasi-energy band-structure of the model in an infinite zigzag nanoribbon geometry, with a finite width. We identify the four integers $C_0^\uparrow, C_0^\downarrow, C_\pi^\uparrow, C_\pi^\downarrow$ (defined later)  that characterize Floquet topological insulators in our model, in each of the phases in Fig.~\ref{figone}
and \ref{figtwo}, by choosing appropriate values of $\omega$, $\alpha$ and $lE_z$.  A representative  diagram for the phase $\mathcal{A}$ has
been shown in Fig.~\ref{figfour} and the remaining diagrams have been relegated to the appendix.
The spectrum has been shown slightly beyond  the `first Floquet-Brillouin zone', 
$-\omega/2<\epsilon_b<\omega/2$, so that the edge states at the zone boundaries are clearly visible.

The first point that we note is the gaps and the edge states at the zone boundaries (at $\epsilon_b = \omega/2 \equiv -\omega/2$).
In the high frequency regime studied earlier, we had  restricted ourselves to frequencies below the zone boundaries (i.e, at $\epsilon=\pm\omega/2$), and hence the edge states at the zone boundary do not appear. However, in this work, our main focus is on the low frequency regime, and one of our aims is to explicitly check that  the Chern number of the band is given by the difference between the number of chiral edge states above and below
the band. How do we count the number of chiral edge states?  As shown in Ref.~[\onlinecite{Rudner2013}], the number of edge modes are related
to the winding number of the Floquet operator.  Unlike the Chern number of a band, which depends only on the stroboscopic
dynamics of the Floquet operator, the winding number has information about the circulation direction,  which gets related to 
the direction of propagation of the edge states.  In a Floquet system, the chirality at a given edge depends on details of the
driving and can be either positive or negative, independent of the chirality of the driving force\cite{Platero2014}. The chirality of the driving force only
provides the required time-reversal breaking.  However, at low frequencies, there is no direct relation between the chirality
of the drive and the chirality of the edge states, since the drive can lead to multiple gap closings and openings with multiple edge states.
Hence, the edge state chirality needs to be explicitly computed for each phase.

%--------------- Fig 5-------------------
\begin{figure}
	\begin{center} 
	\includegraphics[width=0.4\textwidth]{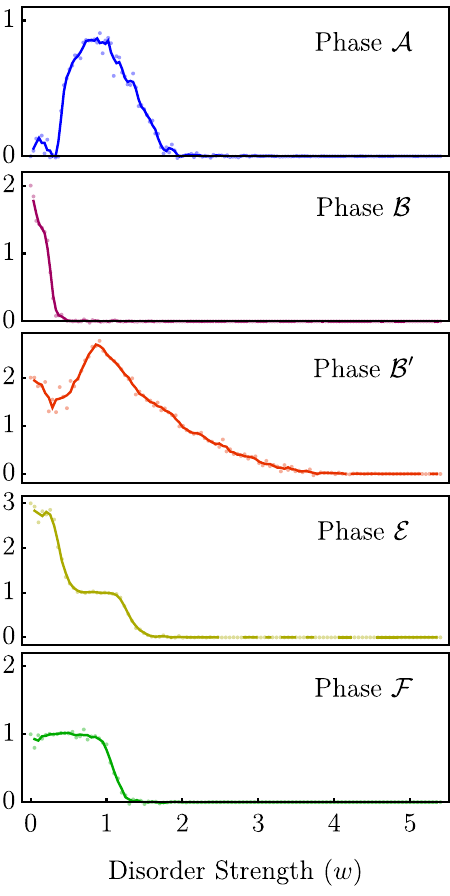}
	\end{center} 
	\caption{The disorder averaged real-space `Chern numbers' for the $\uparrow$ spin sector is shown as a function of the disorder strength $w$ for the phases $\mathcal{A} \dots \mathcal{G}$  depicted in Fig.\ref{figone}. 
	Static uniform disorder is included in the system as an on-site potential. Numerical calculations are  carried out for a 
	24 $\times$ 24 lattice  with open boundary conditions. The Chern number is averaged over 100 disorder configurations and
	the following parameter values ($\alpha$, $\omega$) - (1.4,2.55), (0.3,4.5), (0.5,2.3), (0.2,2.5), (0.9,3.5), (1.6,5.0) were used for  the phases $\mathcal{A}$, $\mathcal{B}$, $\mathcal{B}'$,$\mathcal{D}$, $\mathcal{E}$ and  $\mathcal{F}$ respectively.
	 ($\mathcal{B}$ and $\mathcal{C}$ as well as $\mathcal{G}$ and $\mathcal{D}$ are the same for $\uparrow$ spin as 
	explained in the text).} 
	\label{figfive}
\end{figure}
%-------------------------------------------------

Let us now focus on the Floquet band structure in the various  different phases.
For illustration, let us confine ourselves to the spin up band.
Let us also confine our attention to the left edge ($L$).  The determination of the chirality of the edge
state as shown on the graph is made by actually checking whether the right-moving state (positive slope)
is at the left edge or at the right edge and similarly whether the left-moving slope (negative slope)
is at the left or right edge. This can be done explicitly since we have numerically obtained all the wave-functions. 
We can now easily count the number of chiral edge states at the band-gap
at zero, and at the band gap at $\omega/2$, in the various plots in the panels in Fig.~\ref{figthree} and in the appendix.  
We choose a convention where a right-moving (positive slope in the energy versus momentum plot) at the left $L$ edge state is
assigned a winding number  or chirality $-1$ and a left moving (negative slope) state is assigned a chirality $+1$.
We then compute  $C^\sigma_0$  by taking it to be $-1/+1$  depending on whether the $L$ state ( or states) in the band-gap at zero frequency  is right-moving or left-moving
and adding up the values.
Similarly, in the band-gap at frequency $\omega/2$, we compute  $C^\sigma_\pi$ by taking  $ -1/+1$ for each right-moving/left-moving state and adding
up the values.
For instance, in  Fig.~\ref{figthree}, for the spin-up band, at zero frequency,  there is a single edge state at the left edge which has
negative slope; thus $C^{\uparrow }_0 = +1$.    At the frequency $\omega/2$ also,   there is a single 
edge state at the left edge with negative  slope, thus $C^{\uparrow}_\pi =  +1$ as well. The Chern number of the $\uparrow$ band  in phase $\mathcal(A)$  was computed 
earlier to be $C^\uparrow =1$ which precisely agrees with $C_0^\uparrow -C_\pi^\uparrow$, as expected from Ref.~[\onlinecite{Rudner2013}].

Using the same method,  $C_0^\sigma$ and $C_\pi^\sigma$ can be computed for each of the phases in Fig.~\ref{figone} and \ref{figtwo} and the results are tabulated in Table 1. Note that, as expected, the Chern number of the band, $C^\sigma = C_0^\sigma - C_\pi^\sigma$ in each case.  Note also  that the phases $\mathcal{A},\mathcal{B}, \mathcal{C},\mathcal{E},\mathcal{F}$ in the table are present in both Figs.~\ref{figone} and \ref{figtwo},  whereas $\mathcal{B'},\mathcal{D}$ and $\mathcal{G}$ occur only  in Fig.\ref{figone} and $\mathcal{H}, \mathcal{I}$ and $\mathcal{J}$  only in Fig.~\ref{figtwo}.

\section{Discussions and conclusions}
In comparison with earlier studies of irradiated graphene, the main difference for spin-orbit coupled materials is the fact that the phase boundaries for the spin $\uparrow$ electrons and the spin $\downarrow$  electrons occur at different points in 
the parameter space. Besides, due to the buckling, an external electric field can be applied which can tune the masses at the $K$ and $K'$ points . This external tuning parameter helps in finding new phases as seen in Fig.~\ref{figtwo}, which do 
not exist in graphene.

We have also studied the robustness of each of the phases in the presence of (uniform) disorder. The disorder in the system is modeled as an on-site chemical potential which is taken from a normal distribution distribution of standard deviation $w$, where $w$ serves as the
strength of the disorder in terms of the hopping parameter $t$. In Fig.~\ref{figfive}, we have plotted the disorder averaged real-space 'Chern numbers' of the various phases in Fig.~\ref{figone}, computed using the coupling matrix approach following Ref.~\cite{coupling}. We note that a number of the topological phases are immune to uniform disorder for a reasonable  range of the disorder strength, and starts degrading only for larger values, whereas a few topological phases immediately change
their character even for a relatively small disorder. For a few of the phases,  the robustness against disorder can be understood in terms of the respective values of the quasi-energy gap in the system, but in certain cases (such as contrasting phase $\mathcal{B}$ and $\mathcal{D}$, 
see Appendix), the robustness against disorder may not be simply related to the quasi-energy gap of the system for each of the phases, which can be compared with the disorder strength $w$ required to change the topological order. This is a surprising outcome and is expected to be related to structure of the time dependent Hamiltonian and is a direction for future study. We also note,  in passing,   
that  the phase  $\mathcal{A}$, characterized by zero value of the topological invariant appears  to attain the Floquet topological Anderson insulator phase~\cite{AFAI1,AFAI2} and exhibits two-lead quantized current at the infinite bias limit~\cite{AFAI3}. Further, the robustness of 
a certain phase also implies that any transport phenomena, such as a sum-ruled quantum Hall conductance~\cite{tami1,tami2,sumrule}, should also be protected and might act as signatures to identify the individual phases. This is of particular importance, because  the lack of knowledge of
the occupation of the bands can be circumvented using signatures of the 
edge states.

The low-frequency analysis in this manuscript focuses on spin-orbit 
coupled materials which are silicene, germanene and stanene in condensed 
matter systems. Although theoretical study of experimentally attainable 
parameter values require detailed study as has been done in graphene 
\cite{LF_graphene} we provide the values which are used in this study 
e.g. the phase $\mathcal{A}$ can be realised in silicene with frequency 
$\omega=2.55\, t$ which belongs to near infra red (INR) in 
electromagnetic spectrum (with hopping parameter, 
$t \sim 1\, eV$); amplitude, $A=1.5$ in units of the inverse of lattice 
constant ($a_0=3.84\, A$); external electric field, $E_z=0.173 \,V/ A$ 
\cite{sil1,sil2}; spin-orbit coupling, $\lambda=0.05\, t=3.5\, meV$ 
\cite{sil_SO}.  We note that it might seem experimentally challenging in 
condensed matter systems, the range of parameter values required to 
realise proposed Floquet topological phases are accessible experimentally 
in a photonic crystal structure \cite{F_photonic, Dis_Titum}. Recently, 
A. 
Quelle et al ~\cite{AFAI} provided a driving protocol to realize 
anomalous 
Floquet-Anderson insulating (AFAI) phase in optical lattices.

\vspace{0.5cm}
\section*{Acknowledgments}
We would like to thank Priyanka Mohan, Udit Khanna  and Dibya Kanti Mukherjee for many useful discussions.
The research of R.S. was supported in part by the INFOSYS scholarship for senior students.

\appendix
\new

\section{Edge states in the ribbon geometry for different phases}

In this appendix, we compute the Floquet band structure in a zigzag nano ribbon in all the different phases
which have been shown in Figs.\ref{figone} and \ref{figtwo} in the main text. In the main text, the band diagram for phase $\mathcal{A}$ was already shown; here we 
show the edge-state spectrum for all the remaining phases. The name of the phase, as well as the values
of $C_0^\sigma$ and $C_\pi^\sigma$ are given in the figure itself. As described in the main text, $C_0^\sigma$ and $C_\pi^\sigma$ are computed by taking it to be $-1/+1$  depending on whether 
the $L$ state (or states) in the appropriate band-gap  is right-moving or left-moving at the left edge of the sample  and adding up the values. Note that it is not
always to  visually determine whether or not the gap exists and in ambiguous cases, we have explicitly mentioned that it is gapped. Note also that in the diagrams
of the phases $\mathcal{G},\mathcal{H} $ and $\mathcal{I}$,  the edge states are isolated from the bulk states at zero energy even though the spectrum is not gapped 
( or has an extremely small gap). Thus the computation of the Chern numbers by counting edge states is more reliable than the bulk computation, which can numerically
fail in the absence of a well-defined gap. \\

%--------------- Fig Appendix -------------------
 \begin{figure*}
 \begin{center}
\includegraphics[width=0.45\textwidth, height=0.28\textwidth]{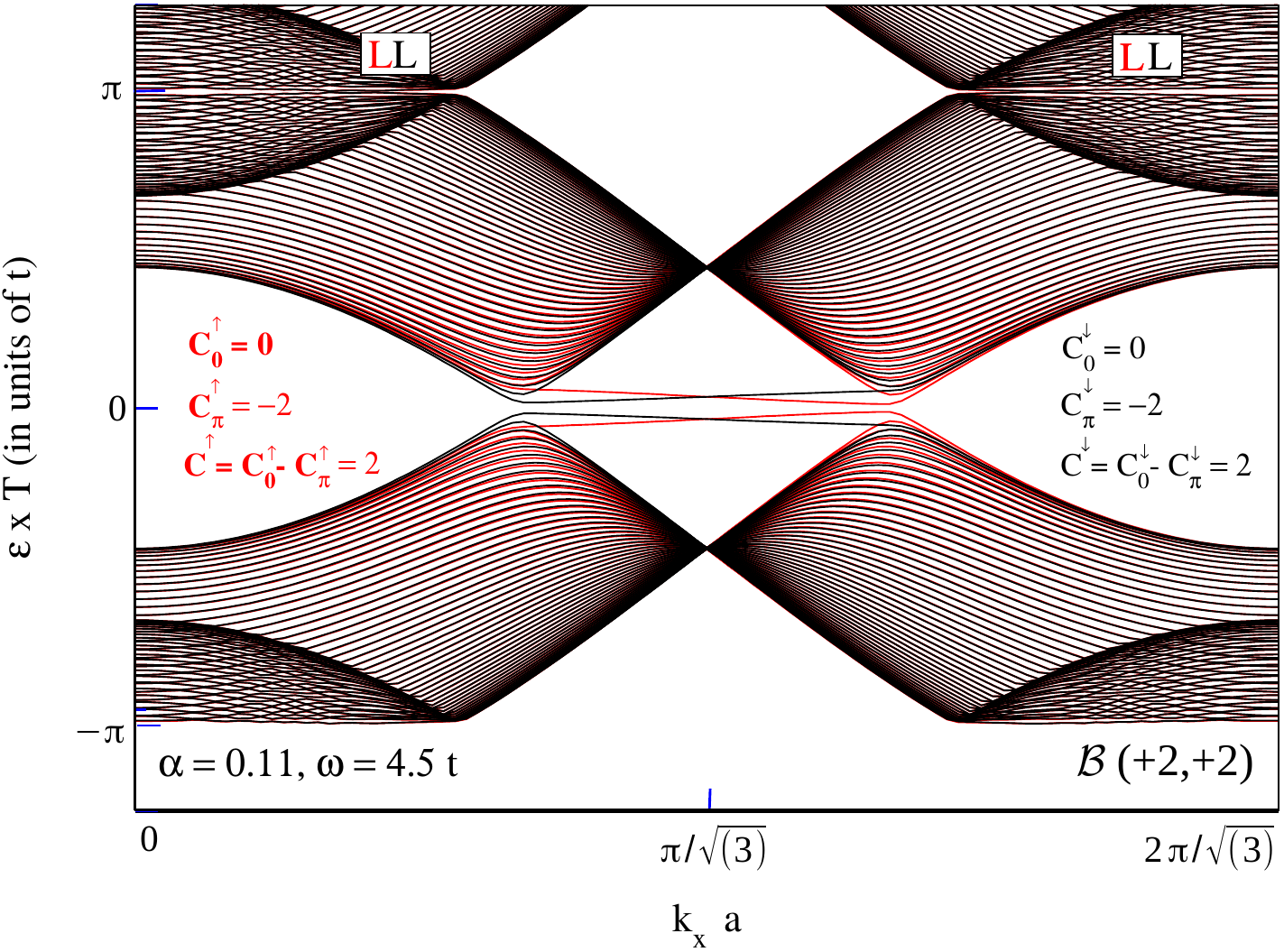}
 \includegraphics[width=0.45\textwidth, height=0.28\textwidth]{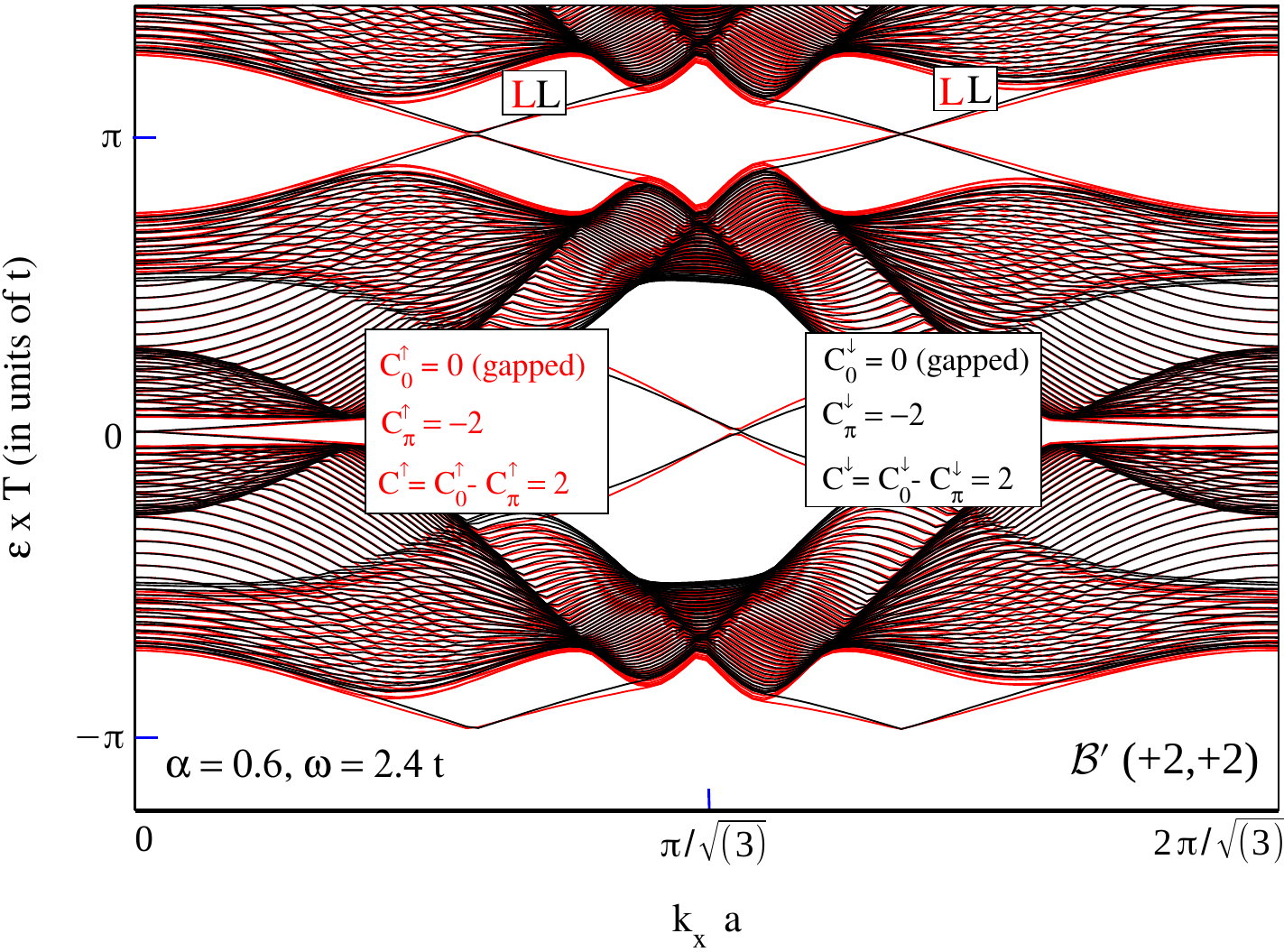}
\includegraphics[width=0.45\textwidth, height=0.28\textwidth]{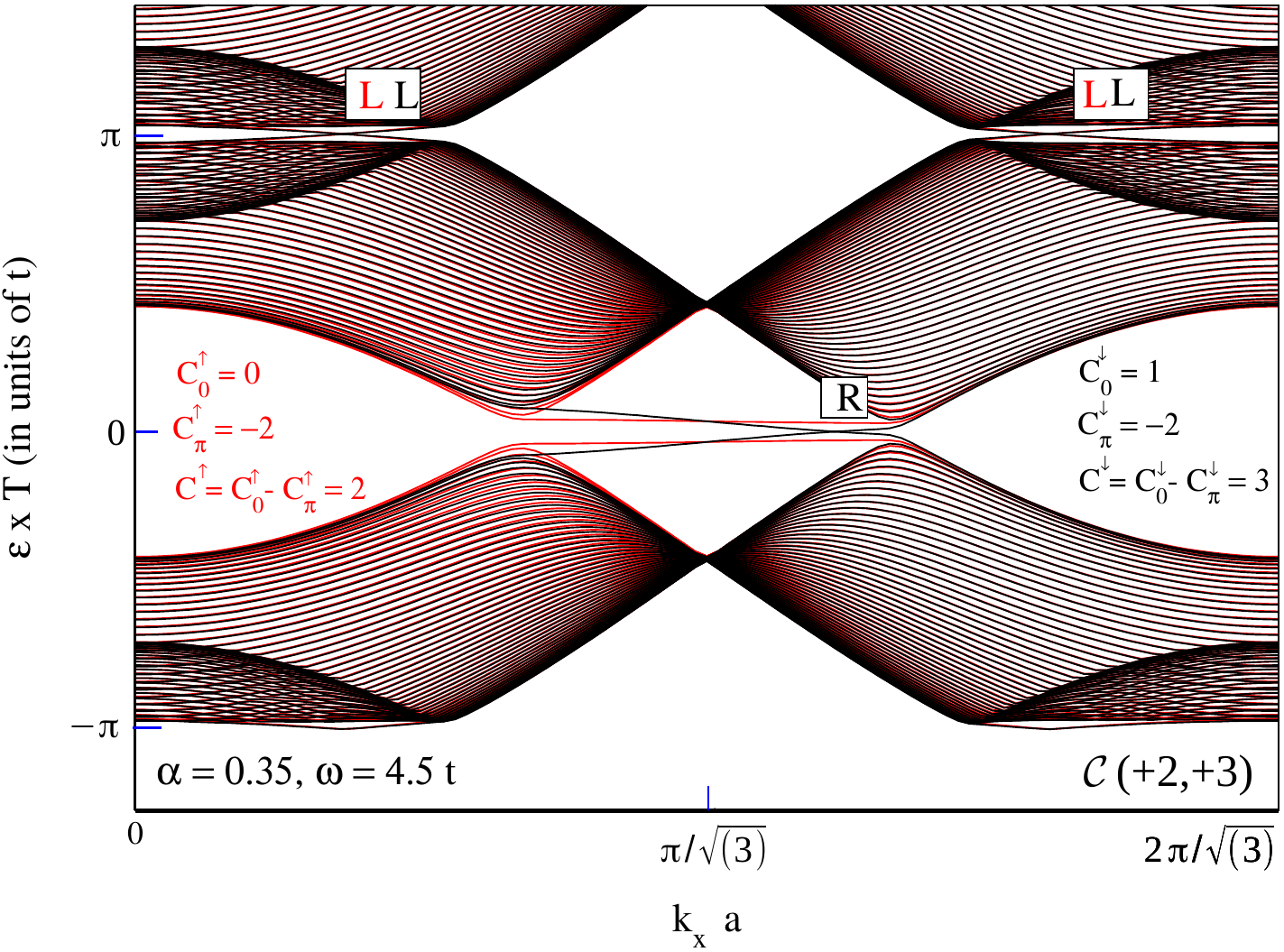}
\includegraphics[width=0.45\textwidth, height=0.28\textwidth]{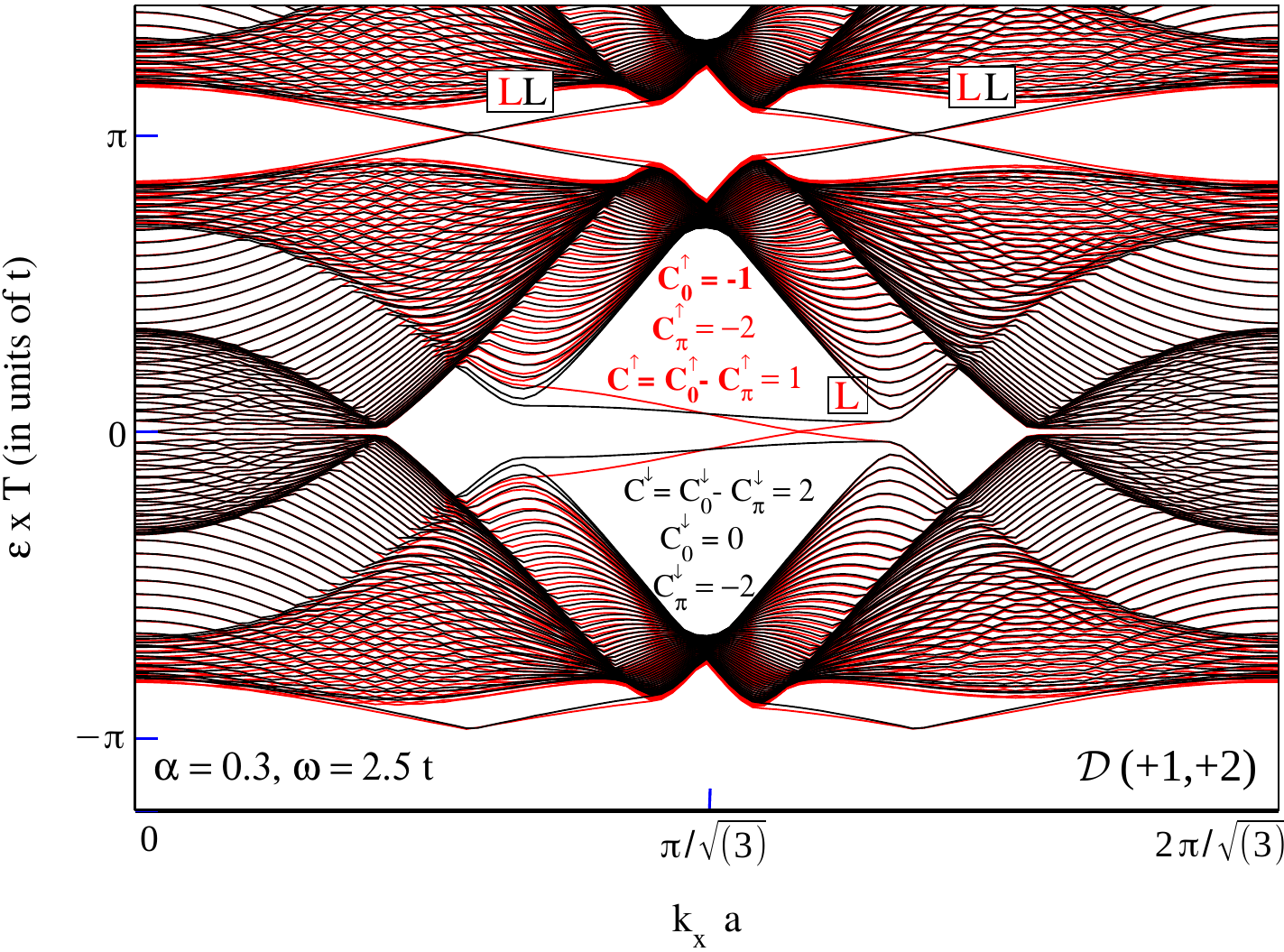}
\includegraphics[width=0.45\textwidth, height=0.28\textwidth]{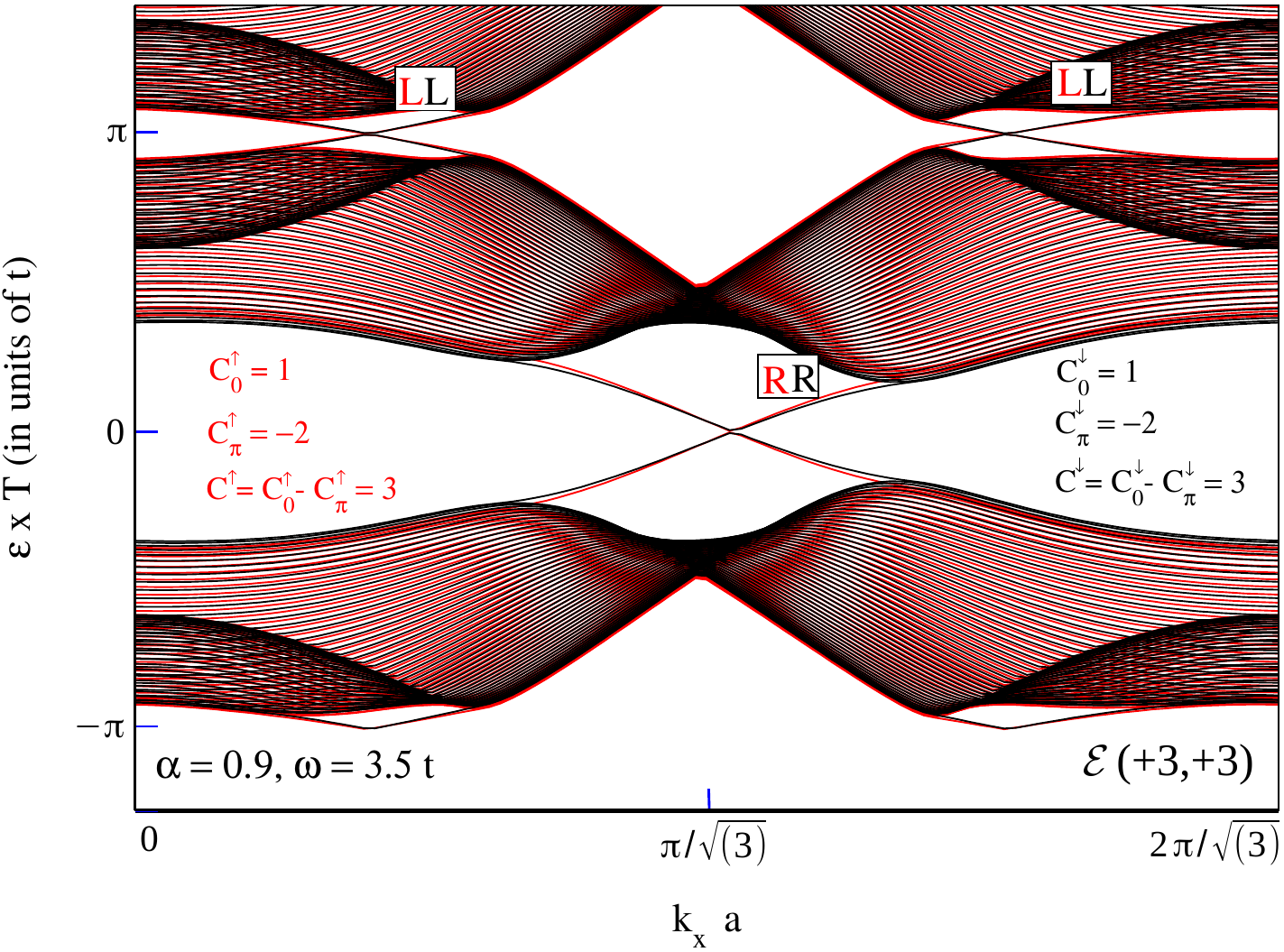}
\includegraphics[width=0.45\textwidth, height=0.28\textwidth]{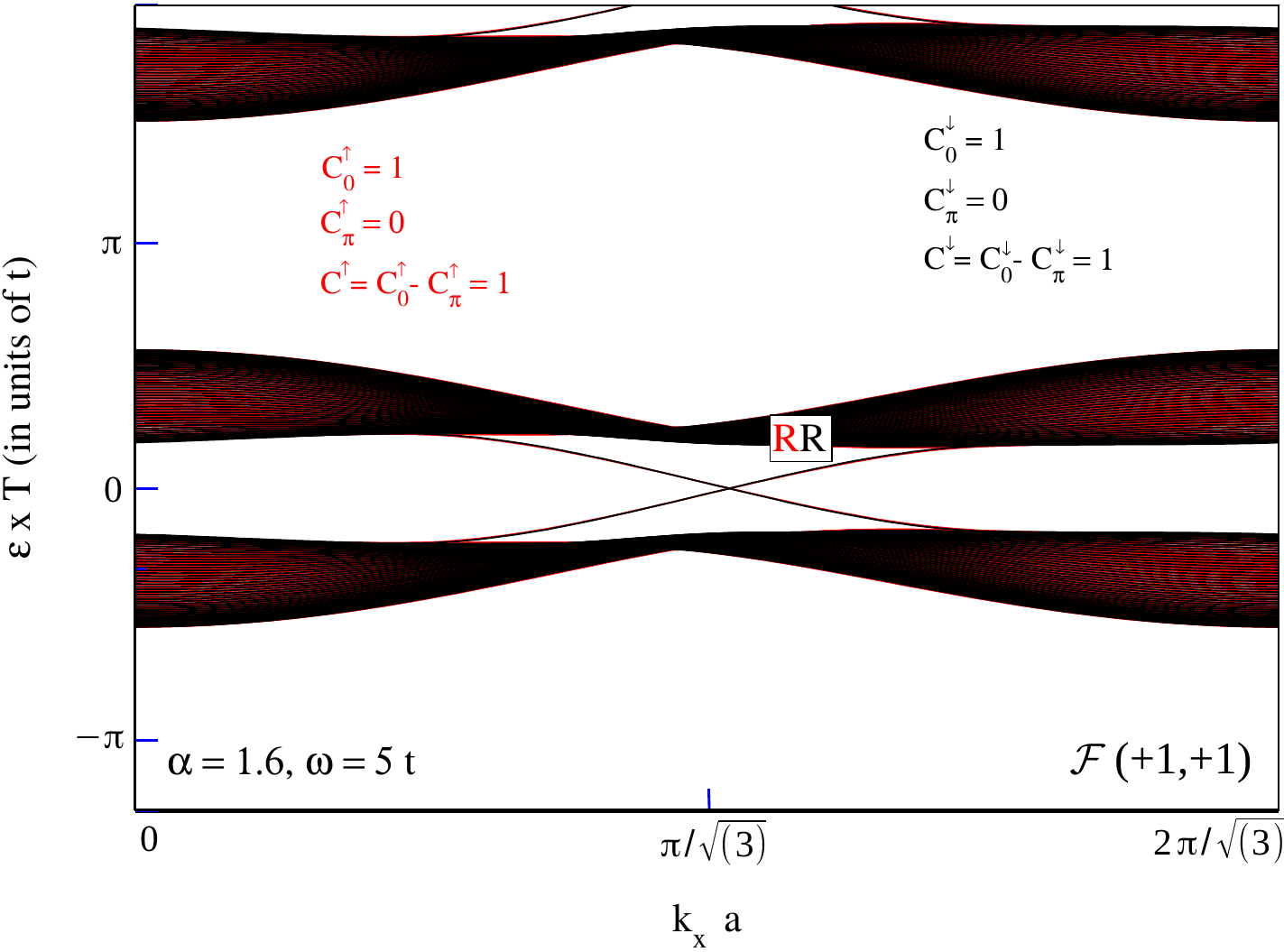}
\includegraphics[width=0.45\textwidth, height=0.28\textwidth]{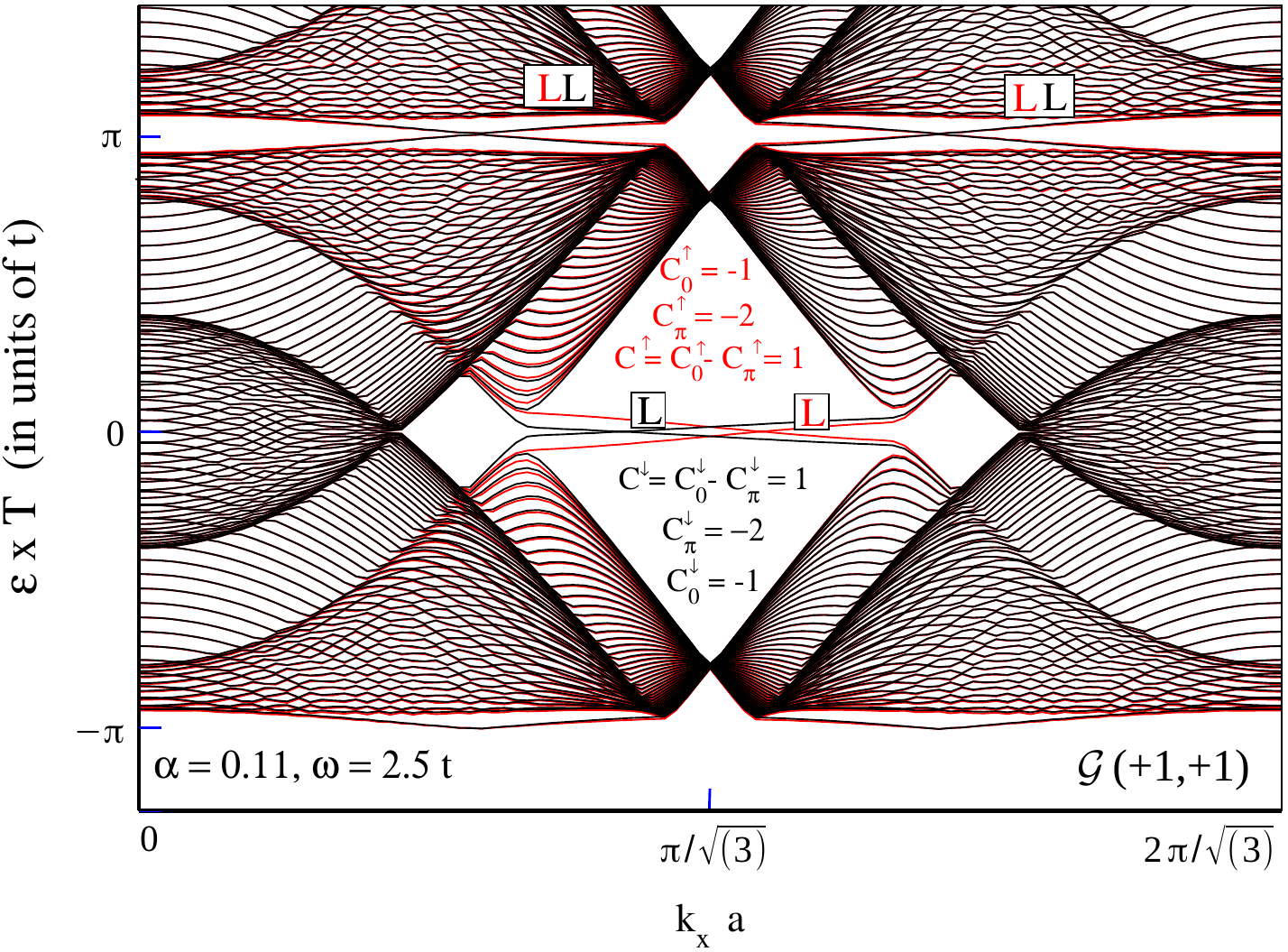}
\includegraphics[width=0.45\textwidth, height=0.28\textwidth]{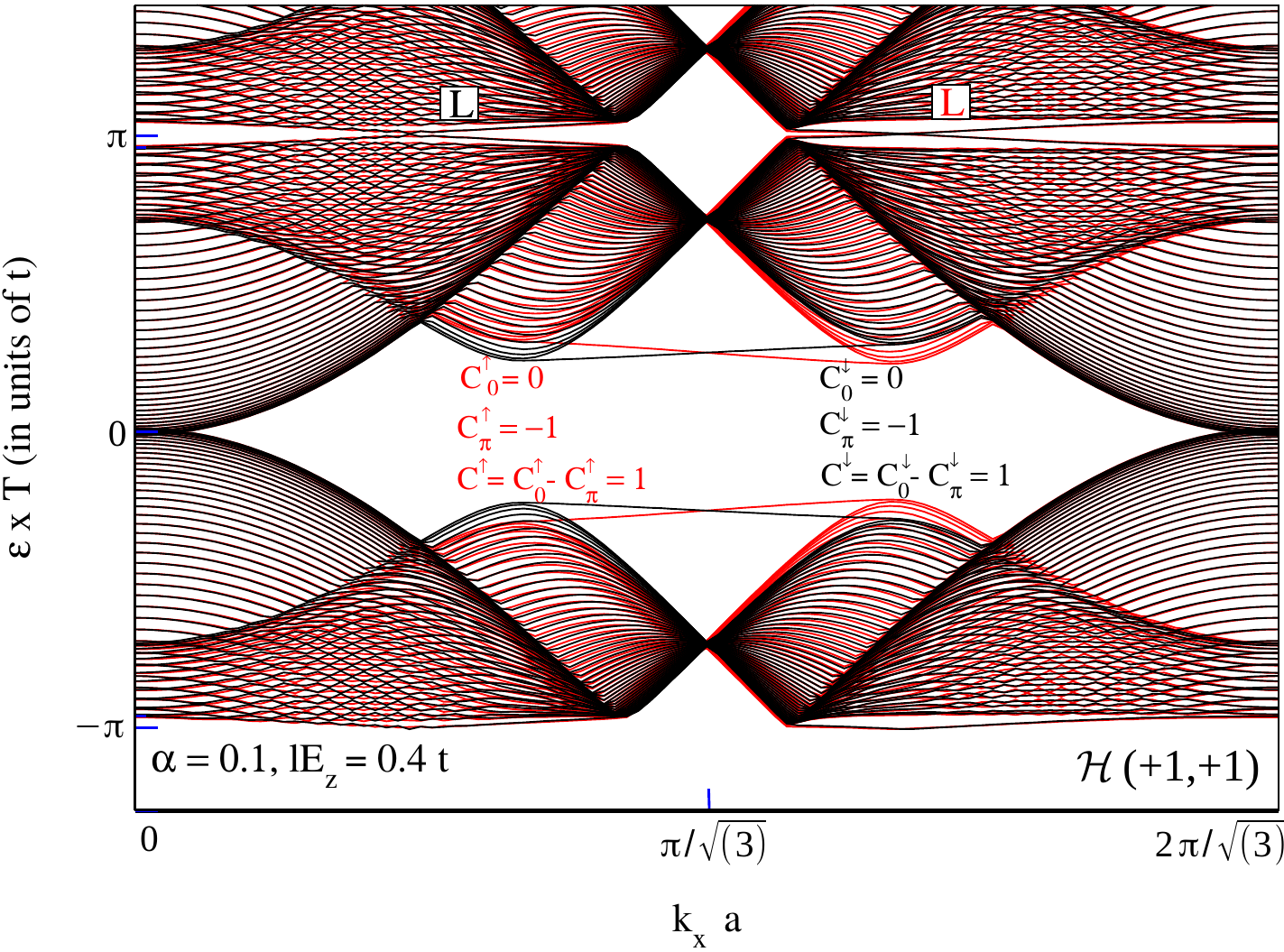}
\includegraphics[width=0.45\textwidth, height=0.28\textwidth]{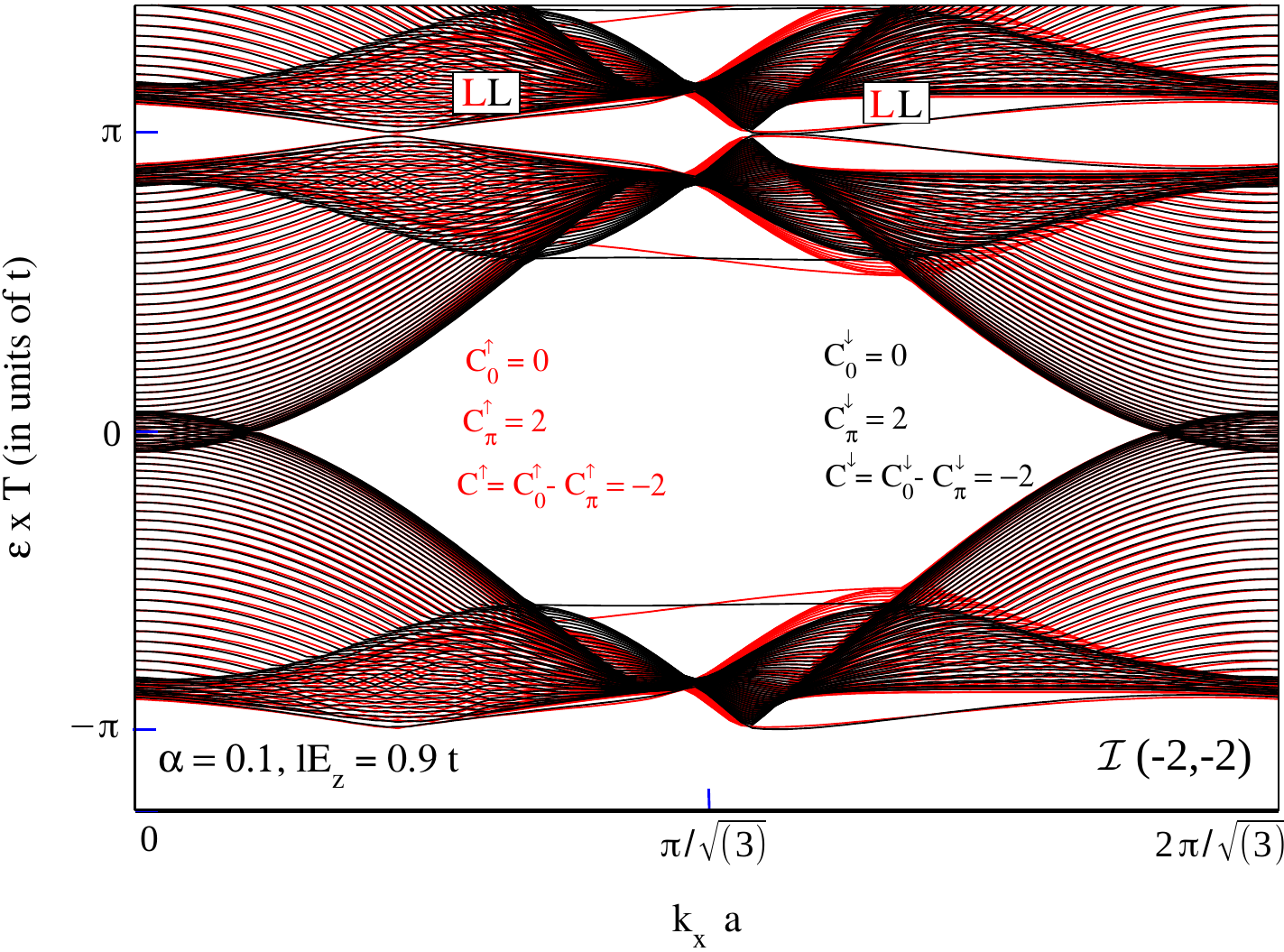}
\includegraphics[width=0.45\textwidth, height=0.28\textwidth]{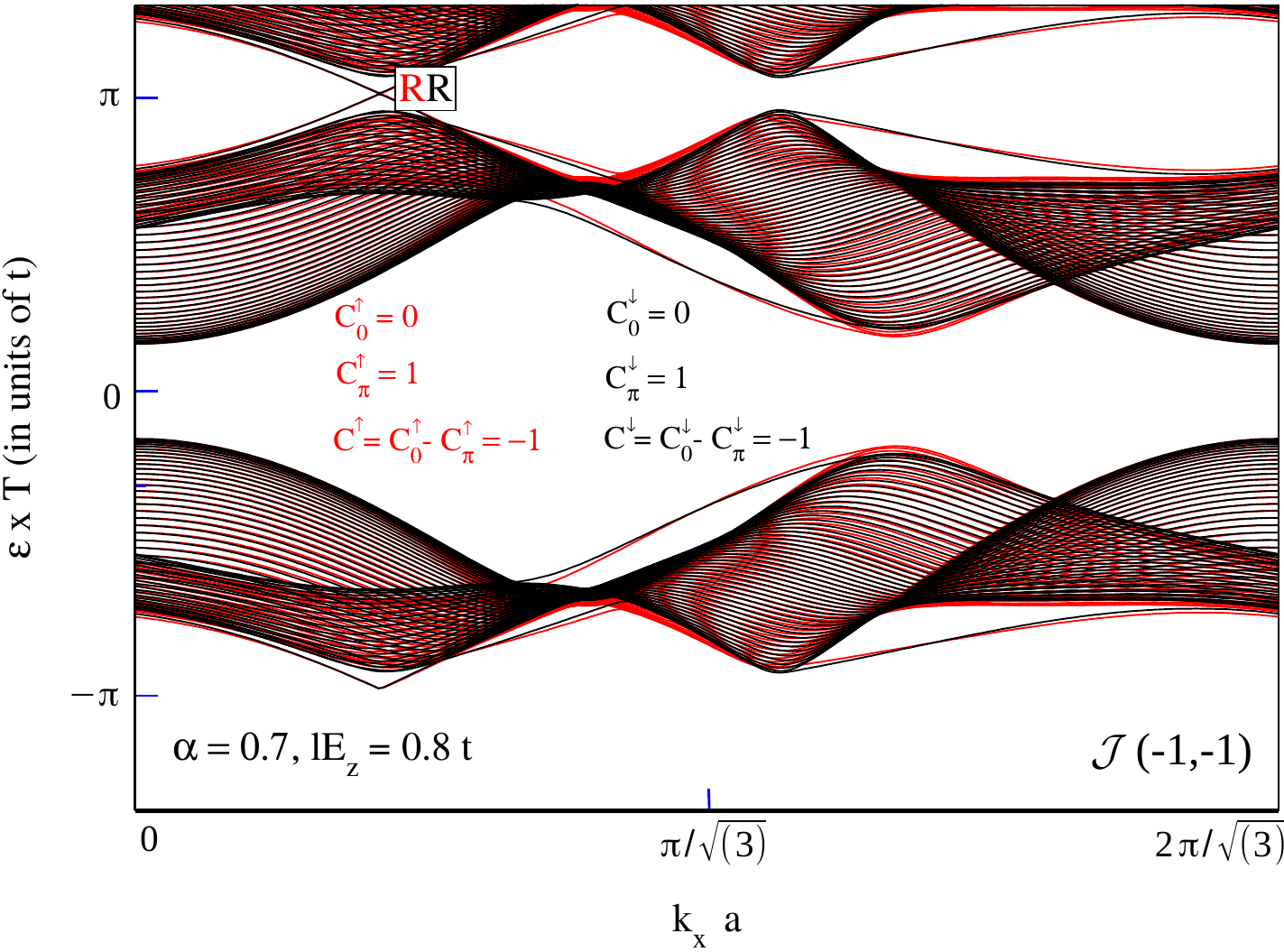}

 \end{center} 
   \caption{} 
   \label{figappendix}
 \end{figure*}
%-----------------------------------------

\end{document}